# Enhancing Multiclass Malware Classification in Resource-Constrained Environments

Abdul Khalek Alve, Alif Rahman, Saadman Zaman, Sazzad Hossen Himel, Muhammad Iqbal Hossain
Department of Computer Science and Engineering
BRAC University, Dhaka, Bangladesh

***Abstract*—The emergence of multi-class malware attacks such as ransomware, spyware, trojans, etc., presents an increasing and serious threat to cybersecurity, particularly in resource-constrained environments like IoT devices. Existing machine learning models have achieved nearly perfect accuracy in binary malware classification but fall short in terms of classifying malware families and individual malware. Additionally, the complexity of these multi-class malware attacks presents a significant challenge of detection in resource-constrained environments, as multi-class detection usually requires high computational capability. This research bridges the gap by enhancing the detection accuracy of multi-class malware classification as well as developing a lightweight model that can run efficiently on resource-constrained devices. In this paper, we propose a robust, lightweight machine learning model featuring LightGBM classifier with SMOTE oversampling and SOM-US undersampling techniques for data balancing, as well as well-engineered feature selection through Genetic Algorithm. The model performed better than the current state-of-the-art models developed on the same dataset in both malware family classification (4 classes) and individual malware type classification (16 classes) with accuracy of 89.1% and 76% respectively. Thus, maintaining a balance between classification accuracy and computational efficiency in resource-constrained environments. Furthermore, we propose another model using Random Forest classifier with an accuracy of 91.2% in malware family classification and 78.7% in individual malware classification. Demonstrating a significant enhancement in terms of accuracy from the current state-of-the-art models.***

***Index Terms*—Machine Learning, Malware Detection, Multi-Class Classification, Random Forest, LightGBM, KNN, Decision Tree, Genetic Algorithm, SMOTE, SOM-US.**

## I. Introduction

The increasing complexity of cyberattacks has driven an explosive increase in the types of malware, ranging from the antiquated virus to the more recent trojans, worms, ransomware, and spyware. These malware programs utilize a broad spectrum of attack vectors and evasion techniques, making them ever more difficult to detect—particularly in real-time and on low-compute platforms such as those utilized in Internet of Things (IoT) networks.

Traditional malware detection has almost been dependent on signature-based methods, which are reactive in nature and lack response to new or polymorphic threats. Machine learning (ML) algorithms, however, have the adaptive strength of learning abstract patterns from data. While ML-based binary classification (malicious or clean) has reached high levels of accuracy, the more complicated multiclass classification task of recognizing a specific malware family or type is underdeveloped, particularly for low-resource platforms.

The drive for this work arises due to the inadequacy of existing multiclass detection methods on embedded systems. Despite the fact that deep models and ensemble methods have a high potential, their computational cost in most instances outweighs the capability of common IoT devices. In addition, most existing models do not address real-world pragmatic issues such as dataset imbalance and redundant features, leading to poor performance in actual deployment.

The CIC-MalMem-2022 dataset has been widely defined as a benchmark for memory-based malware detection. The dataset contains a full set of features extracted from malicious and clean system states in memory, offering a robust base for behavioral inspection. While numerous studies have achieved high accuracy binary classification on this dataset, multiclass performance remains low due to the employment of obfuscation techniques and the significant resource cost of deep learning models. The best performance in family-level classification is currently 87.1%, while individual malware classification accuracy remains 72.6%.

This paper introduces a high-performance and resource-conserving malware detection system that is appropriate for deployment in environments with limited resources. The aim of the research is two-fold, i.e., to improve the classification accuracy at malware family and individual type levels, and to minimize the computational effort required for detection. By utilizing effective data balancing techniques, best feature extraction, and lightweight classifiers such as Random Forest and LightGBM, the presented approach addresses significant limitations of present approaches meant for embedded and IoT-based systems.

## II. Related Works

Shafin et al. [1] (2023) proposed a lightweight multi-class malware detection model based on the feature-learning capabilities of Convolutional Neural Networks (CNN) with bidirectional Long Short-term Memory (LSTM). The hybrid model called by authors as CompactCBL is designed for fast processing with a compact size, making it compatible with IoT devices. They also built a similar but larger model called RobustCBL with the same OMM dataset named 'CIC-Malmem-2022' [1]. Experimental results show that CompactCBL achieved 99.92% accuracy for Binary classification, 84.22% for family classification and 71.42% for individual attack detection. However, it is outperformed by the RobustCBL model which achieved 99.96%, 84.56% and 72.6% in Binary, family and individual classifications respectively [1]. Since both of the models underperformed in terms of individual attack type detection, there is need for further research to enhance the accuracy in this regard. Moreover, there is room for improvement in malware detection, especially for unknown or zero-day attacks by incorporating semi-supervised or unsupervised learning models to increase detection capabilities while maintaining the model's suitability with IoT devices in resource constrained environments [1] .

Shhadat et al. [2] (2020), researchers used machine learning mechanisms to enhance the accuracy rate of hidden malware detection and classifying those malwares. Machine learning models have been used on a benchmark dataset such as Random Forest, Support Vector Machine, Decision Tree, Logistic Regression, Na¨ıve Bayes, and Adaptive Boost. RF classifier is a methodology which is used for selecting features as well as cross validation to split data. In these methods they investigated the relevant features and rejected unnecessary features which increased the efficiency and accuracy. In a tree based data lowest impurity is found at the end where cutting specific nodes they found the subset of features. In this paper the imbalance dataset problem has been mitigated by 15-fold cross-validation. The experiment results in maximum accuracy for binary classification by (98.2%) using decision tree and (95.8%) for multi-class classification by using Random Forest but for the Bernoulli Na¨ıve Bayes the accuracy rate for binary classification is (91%) and (81.8%) for multi-class classification. In this research, they faced challenges with obfuscated malware and they have suggested enhancing real time detection abilities to integrate more diverse datasets.

Abualhaj et al. [3] (2025) introduced a novel malware forensic investigation framework integrating the Gray Wolf Optimization (GWO) algorithm with various machine learning classifiers to increase malware detection precision using memory dump data. The primary problem addressed is that malware detection is difficult especially across multiple classes due to high-dimensional features, redundancy of data, and noise, increasing false positives as well as computational cost. To overcome this, the authors constructed a Digital Forensics Investigation (DFI) model that uses GWO for best feature selection with a significant decrease in the dimension of the dataset without sacrificing valuable features. The main contribution is to boost the accuracy and efficiency of malware classification, particularly in the context of memory-resident malware. The research emphasizes the balanced CIC-MalMem-2022 dataset with 58,596 memory dump entries and 55 features of 16 families of malware. Preprocessing was conducted on data using label encoding and Min-Max normalization. GWO was then utilized to select the optimal features, reducing them to 4, 8, and 16 features for binary, 4-class, and 16-class classification, respectively. Different classifiers were tried out, including Random Forest (RF), Decision Tree (DT), Support Vector Machine (SVM), Naive Bayes (NB), and K-Nearest Neighbors (KNN). RF was the best among them with 75.6% in 16-class classification, 86.34% in 4-class, and 99.93% in binary classification superior to the state of the art on the same set. The model is highly robust, scalable, and has improved detection performance, especially in binary scenarios. Subsequent studies will entail evaluating hybrid optimization techniques and testing the model on diversified data sets to improve generalizability and operational effectiveness in real forensic environments.

Dener et al. [4] (2022) incorporated big data techniques to enhance scalability in malware detection by utilizing frameworks like Apache Spark and PySpark. In the binary classification, they have applied various ML and DL models, of which logistic regression achieved the highest detection accuracy at 99.97%, followed by 99.94% in the Gradient Boosted Tree and 98.41% in the Naive Bayes model [4]. These findings align with the earlier studies which showed the effectiveness of DL and ML approaches in malware detection using big data. In a nutshell, the memory analysis together with advanced ML or DL models shows great potential in enhancing malware detection accuracy in cybersecurity frameworks.

S. S. P et al. [5] (2023) in their study focused on detecting and classifying multiple types of obfuscated malware (ransomware, spyware, and trojan) using machine learning techniques. The research implemented Random Forest, Decision Tree and Gradient Boosting models. The CIC MalMem 2022 dataset was used for this research. It contained memory analysis data of malware samples. These were used to do both binary and multi-class classification. Feature engineering was applied during preprocessing. The Random Forest model achieved the highest accuracy for multi-class classification at 89.07% accuracy. Meanwhile all the models were successful in binary classification. It should be notable that the model outperformed convolutional neural network models of that time. According to the authors, future improvements should focus on getting more accuracy in multi-class classification and explore hyperparameter tuning as that would help tweak the model to get more optimal outcomes.

In this paper by Carrier et al. [6] (2023) proposed a lightweight multi-class malware detection model based on the feature-learning capabilities of Convolutional Neural Networks (CNN) with bidirectional Long Short-Term Memory (LSTM). The hybrid model called by authors as CompactCBL is designed for fast processing with a compact size, making it compatible with IoT devices. They also built a similar but larger model called RobustCBL with the same OMM dataset named 'CIC-Malmem-2022'. Experimental results show that CompactCBL achieved 99.92% accuracy for Binary classification, 84.22% for family classification, and 71.42% for individual attack detection. However, it is outperformed by the RobustCBL model which achieved 99.96%, 84.56% and 72.6% in Binary, family and individual classifications respectively . Since both of the models underperformed in terms of individual attack type detection, there is need for further research to enhance the accuracy in this regard. Moreover, there is room for improvement in malware de- tection, es-pecially for unknown or zero-day attacks by incorporating semi-supervised or unsupervised learning models to increase detection capabilities while maintaining the model's suitability with IoT devices in resource constrained environments.

Hamidouche et al. [7] (2024) it addresses the issue regarding the challenge of implementing effective real-time threat detection on resource constrained devices such as IoT devices. IoT and other such devices lack computational power and memory to support advanced security systems like intrusion detection systems or malware protection making them an easy target for cyber threats. The main objective is to develop a lightweight DNS tunneling detection model that can used by resource constrained devices the evaluate the feasibility of implementing ML and DL methods for real time detection and then show the effectiveness of the model developed.The methodology followed is Data Collection, feature extraction, model training and performance evaluation is done. The paper uses several ML models like decision tree, random forest, KNN, SVM and DNN but the random forest model had the highest accuracy. While the model could achieve high accuracy in a trained environment, ensuring the model could work efficiently without delay in a real time environment. Developing a more lightweight ML model for higher accuracy with low latency.

Chen and Ren [8] (2023), the research was based on the malware family classification where multi-features fusion has been used and it is an efficient boosting-based technique. The researchers used a dataset of Microsoft Malware Classification Challenge (BIG 2015). Experimenters applied a forward feature mechanism for step by step selection to mix features of plausible binary malware with assembly malware to create new features. Support Vector Machine (SVM), (XGBoost), Random Forest (RF), (KNN), and adaptive boosting (AdaBoost), are some machine learning techniques they have used for fusion feature set and classifications. They have applied algorithms such as tree-boosting-based LightGBM, XGBoost, and CatBoost that the accuracy results they have found is (99.87%), (99.84%) and (99.76%) respectively. Researchers suggested continuing to work with emphasis striking a balance between time and accuracy.

Kumar [9] (2024) introduced a novel under-sampling technique known as SOM-US, leveraging self-organizing maps (SOM) to address the class imbalance problem prevalent in many real-world datasets, particularly for software defect prediction. This method stands out by clustering the majority class to reduce its samples effectively, ensuring a more balanced dataset for training classifiers. The approach was tested using logistic regression on a NASA software defect dataset, where it demonstrated significant improvements over six existing under-sampling techniques across performance measures like G-measure and nMCC. Class imbalance is a major challenge in machine learning, leading to biased classification models that favor the majority class. Traditional under-sampling techniques, such as random under-sampling (RUS) and Tomek Links (TL), remove majority class instances to balance the dataset but risk losing important information. Kumar's study is distinct in applying a neural network technique to undersampling, contrasting with more conventional methods that do not integrate advanced clustering processes. The research outcomes suggest that SOM-US notably enhances the predictive accuracy of models dealing with class imbalances by focusing on the majority class's structural nuances rather than merely reducing its size. This method could potentially be extended to other domains facing similar challenges, such as fraud detection and medical diagnosis, where class imbalance can significantly skew predictive performance. It proves a promising direction for future research, especially in exploring the adaptability of this technique across various application areas and larger datasets.

Saaidah et al. [10] (2024) mainly focuses on improving the malware detection made by the KNN machine learning algorithm using the firefly optimization algorithm (FOA). The study uses the MalMem-2022 dataset for binary as well as multiclass classification. The focus is mainly on improving the feature selection using the firefly algorithm. The dataset preprocessing is done by applying label encoding and min max scaling, then FOA is applied to remove all redundant and irrelevant features. The firefly algorithm selects 14 features for binary classification, 26 features for 4-class multiclass classification(Family), and 29 features for 16-class multiclass classification(Individual) out of the 55 total given features. Once feature selection is completed, KNN algorithm is applied. The results show that while FOA improves multiclass classification, it does not enhance binary classification performance. KNN without FOA achieves 99.97% accuracy in binary classification but with using FOA, accuracy slightly drops by 0.04%. However, in Family class classification the FOA improves the accuracy from 82.22% to 83.74% which is an increase of 1.52% and in Individual class classification the accuracy increases from 66.94% to 69.58% which is an increase of 2.64%. This suggests that while FOA may not be beneficial for binary classification, it enhances accuracy in multiclass scenarios. Even after improving slight accuracy, the study has many limitations. Lack of class balancing could lead to biased results, favoring the majority classes. Additionally, the paper does not compare FOA enhanced KNN with other machine learning models like Random Forest or LightBGM making it unclear if FOA actually performs better. Another concern would be overfitting and also it does not compare its performance with other feature selection processes available. In conclusion, the study demonstrates that FOA can improve KNN's performance in multiclass malware detection, enhancing accuracy. However, issues like class imbalance and lack of model comparisons limit its application in a real life scenario. Future research should explore alternative Models, class balancing techniques, and FOA's computational cost to provide a more comprehensive understanding of its role in malware detection and cybersecurity.

In the paper Talukder et al. [11] (2022) the main aim is to propose a hybrid method combining the machine learning and deep learning methods to improve the current detection accuracy. Network Intrusion Detection Systems (NIDS) play an important role in guaranteeing network security, but present models struggle with accuracy and dependability due to the enormous amount of data they must handle [11]. In the proposed methodology they used SMOTE (Synthetic Minority Over-sampling Technique) to handle the imbalance in the data in datasets. For feature detection techniques they employed the XGBoost for selecting the most relevant features by reducing dimensionality and improving the computational efficiency [11]. The study includes numerous machine learning and deep learning techniques, such as Random Forest (RF), K-Nearest Neighbor (KNN), Multilayer Perceptron (MLP), Decision Tree (DT), Artificial Neural Network (ANN) and also Convolutional Neural Network (CNN) [11]. These were tested on the KDDCUP99 dataset as well as the CIC-MalMem-2022 dataset. The performance of these models was assessed using metrics such as accuracy, precision, recall, F1-score, AUC score, ROC curve, Mean Absolute Error (MAE), Mean Squared Error (MSE), and Root Mean Squared Error (RMSE) [11]. The results achieved are very high accuracy on CIC-MalMem-2022 and on KDDCUP99 with no over-fitting or Type-1 and Type-2 issues . In conclusion the model detected very high accuracy exceeding any existing models. The hybrid approach of using ML and DL along with efficient processing of data and feature selection has a big role to play in this high accuracy and efficiency success.

Cassel and Majd [12] (2024) further worked on 'CIC Malmem-2022' OMM dataset to address the challenges in detecting multi-class obfuscated malware in resource constrained environments, especially in family attack classification. Prior to their research, the detection systems primarily focused on binary classification except for few CNN based multi-class classification deep learning models such as Shafin et al [1]. Those deep learning models are less suitable for resource-constrained IoT devices because of their complexity and high resource consumption. To overcome this, the authors proposed a lightweight machine learning model to detect multi-class obfuscated malwares more accurately. The approach combines SMOTE (Synthetic Minority Over-sampling Technique) for generating additional samples of minority classes with Tomek Links algorithm to undersample the majority class to enhance model performance. The hybrid model coupled with augmented data helps address class imbalance which is a common barrier in malware detection. Using these techniques, the authors trained a Random Forest model on the dataset. Subsequently, experimental results demonstrated that the model achieved 87.1% accuracy in detecting malwares at family level, outperforming the previous state of the solution, RobustCBL (84.56% accuracy). In addition, this lightweight model is suitable for deployment in IoT devices, providing an effective solution in resource constrained environments. In short, the paper highlights the potential of using hybrid data augmentation techniques to improve multi-class malware detection.

Hossain and Islam [13] (2024) proposed a machine learning framework that enhances detection capabilities for both binary and multi-class obfuscated malware through comprehensive data preprocessing techniques based on 'CIC Malmem-2022' OMM dataset. These techniques include memory dump normalization, categorical data encoding and SMOTE to address class imbalance. Additionally, feature selection methods such as Chi-Square tests and mutual information analysis refine the focus of the model on key indicators of obfuscated malware. Central to their approach is the Ensemble-based Classifier, chosen for its robustness in handling complex data structures. The results show that the proposed model achieves an impressive detection accuracy exceeding 99%, significantly outperforming existing models in the field. But they have not made it lightweight and their model is most probably not compatible with resource constrained environments since the experimental setup is conducted on a high-performance personal computer.

Al-Qudah et al. [14] (2023) proposed an effective method of one class classifier for detecting malwares of memory dump. In this experiment a dataset has been used named "MalMem-Analysis2022" which is balanced, tested for detecting obfuscated malware. Two approaches were used where the OCC-PCA models were used for reducing the number of features where OCC means one class classifier as well as PCA means Principal Component Analysis and the TOCC method used for retaining the reduced features. In these two approaches, the OCC-PCA model achieves (99.4%), TNR achieves (99.3%) accuracy rate, respectively. The paper solution focused on detecting memory dump malware accurately. Improving the model's robustness against obfuscated techniques, incorporating more sophisticated features and testing the model's performance on larger dataset will enhance generalizability.

Louk and Tama [15] (2022) in their paper focused on PE (portable executables) malwares using tree-based ensemble models. They conducted the analysis on three datasets (BODMAS, Kaggle, CIC-MalMem-2022). The models that were used are Random Forest, XGBoost, CatBoost GBM and lightGBM. They fine-tuned the models and used hyperparameters to optimize the models. After the training and testing process, they evaluated their findings with existing studies on the same datasets using the k cross-validation technique. For the Kaggle dataset, GBM(H2O) performed the best. While the XGBoost(native) model performed the best for the BODMAS dataset. Random Forest in particular had 100% accuracy, precision and F1 score for the CIC-Malmem-2022 dataset. The results showed that their preprocessing and modifications to the models allowed them to have higher scores on all fronts. Future improvements as stated by the authors were to explore the underexplored domain of deep neural networks regarding tabular data as well as expand on the explainability of tree-based ensemble models.

Vidhya and Srivastava [16] (2023) in their paper aimed to do malware classification and divide them into nine different families using ML models. The paper also mentioned endpoint security and the need to improve the quality of it. The researchers used Microsoft BIG dataset, which includes files of nine different malware families. The byte files were then converted into feature datasets using normalization with unigram features. Random Forest, LightGBM, and XGBoost models were used for the classification. Log loss was the criteria used to evaluate the models. XGBoost outperformed the other models, getting the lowest log loss value of 0.0467. Therefore, XGBoost was found to be the best among the models in the case of multiclass classification. Even though this process doesn't use the confusion matrix, false positives and false negatives are penalized in log loss. Hence the importance of reducing false negatives were emphasized. The authors suggested future works should use deep learning models like Convoluted Neural Networks and Long Short-Term Memory. Also explore GANs which is a type of adversarial technique, to enhance multiclass classification.

Bau et al. [17] (2024) In their paper aimed to evaluate malware detection focusing on multiclass malware classification and use of both static and dynamic features. They also focused on malware that harms android devices. Using the machine learning models Random Forest, Artificial Neural Network and Convolutional Neural Network, they conducted the research. Static classifications focus on permissions and intents meanwhile dynamic classification focuses on API calls and network flows. Through the training and testing process, it was found out that Random Forest model outperformed both the Artificial Neural Network and the Convolutional Neural Network, getting the highest accuracy of 95.30% in static analysis and 78.79% in Dynamic analysis. Dynamic analysis across the board had lower accuracy than static. It should be noted that they did not focus on reducing false positives specifically, just increase the accuracy in general. We want to focus on that more as stated in our objectives. The authors concluded that Random Forest is a more effective model for multiclass classification when we consider less resource consumption because of its efficiency. Future research scopes mentioned where to focus on hybrid methods to combine static and dynamic analysis for better outcomes.

Altınkaya et al. [18] (2025) proposed a new privacy-protecting framework known as FEDetect, utilizing federated learning to accomplish malware detection and categorization effectively with the guarantee of data privacy. The problem here in the paper is traditional centralized malware detection systems, which pose a tremendous threat to the secrecy of data due to the need for the central server to collect personal data of users. Their aim is to alleviate such privacy concerns by decentralizing model training on end-user devices through federated learning (FL) so that raw data stays local while deriving the advantages of collective learning. To solve this issue, the authors developed and emulated a federated learning system using Feedforward Neural Networks (FNN) and Long Short-Term Memory (LSTM) models. Their approach was to develop 22 models, federated and non federated, and compare their performance. The system mimics local model training on emulated user devices and utilizes the FedAvg algorithm with the Adam optimizer for aggregating local models into a global model. The process preserves privacy while retaining high accuracy.The dataset used was the CIC-MalMem-2022, with 58,296 instances of four classes: benign, ransomware, spyware, and Trojan Horse. Preprocessing was conducted by standardization and splitting for binary and multiclass classification tasks.The key contributions are to achieve 99.9% binary classification accuracy and 84.5% multiclass accuracy, demonstrating that FL performs as well as traditional methods but with the additional security of significantly enhanced data security. The paper also provides extensive pseudo-code for simulation and observation on scalability with various numbers of users (8 to 128), so it is well worth considering to implement FL in real-world cybersecurity applications.

## III. Dataset Analysis and Preprocessing

### A. Dataset Collection

| Index | Feature Name | Index | Feature Name |
|---|---|---|---|
| 1 | pslist.nproc | 29 | malfind.protection |
| 2 | pslist.nppid | 30 | malfind.uniqueInjections |
| 3 | pslist.avg_threads | 31 | psxview.not_in_pslist |
| 4 | pslist.nprocs64bit | 32 | psxview.not_in_eprocess_pool |
| 5 | pslist.avg_handlers | 33 | psxview.not_in_ethread_pool |
| 6 | dlllist.ndlls | 34 | psxview.not_in_pspcid_list |
| 7 | dlllist.avg_dlls_per_proc | 35 | psxview.not_in_csrss_handles |
| 8 | handles.nhandles | 36 | psxview.not_in_session |
| 9 | handles.avg_handles_per_proc | 37 | psxview.not_in_deskthrd |
| 10 | handles.nport | 38 | psxview.not_in_pslist_false_avg |
| 11 | handles.nfile | 39 | psxview.not_in_eprocess_pool_false_avg |
| 12 | handles.nevent | 40 | psxview.not_in_ethread_pool_false_avg |
| 13 | handles.ndesktop | 41 | psxview.not_in_pspcid_list_false_avg |
| 14 | handles.nkey | 42 | psxview.not_in_csrss_handles_false_avg |
| 15 | handles.nthread | 43 | psxview.not_in_session_false_avg |
| 16 | handles.ndirectory | 44 | psxview.not_in_deskthrd_false_avg |
| 17 | handles.nsemaphore | 45 | modules.nmodules |
| 18 | handles.ntimer | 46 | svcscan.nservices |
| 19 | handles.nsection | 47 | svcscan.kernel_drivers |
| 20 | handles.nmutant | 48 | svcscan.fs_drivers |
| 21 | ldrmodules.not_in_load | 49 | svcscan.process_services |
| 22 | ldrmodules.not_in_init | 50 | svcscan.shared_process_services |
| 23 | ldrmodules.not_in_mem | 51 | svcscan.interactive_process_services |
| 24 | ldrmodules.not_in_load_avg | 52 | svcscan.nactive |
| 25 | ldrmodules.not_in_init_avg | 53 | callbacks.ncallbacks |
| 26 | ldrmodules.not_in_mem_avg | 54 | callbacks.nanonymous |
| 27 | malfind.ninjections | 55 | callbacks.ngeneric |
| 28 | malfind.commitCharge | | |

Fig. 1: Dataset Features

The "CIC-MalMem-2022" dataset is a crucial resource for the study of multiple types of malware and their corresponding memory analysis, developed by the Canadian Institute for Cybersecurity. Created in 2022, this dataset provides a holistic collection of memory dumps from systems infected with those malware as seen in Figure 1. This resource is designed to support the development and testing of advanced machine learning models for malware detection and analysis. The dataset contains detailed metadata for each memory sample, such as the type of malware, the attack vector, and system impacts. It delivers a real-world testbed for developing robust detection algorithms capable of identifying and mitigating threats based on memory analysis, which enhances its value for researchers and cybersecurity professionals. The "CIC-MalMem-2022" dataset is crucial not only because of its size but also for its focus on the dynamic aspects of malware behavior in system memory, making it a vital tool for improving cybersecurity defenses.

### B. Data Cleaning

First, the 'CIC MalMem-2022' dataset has been loaded into Google's Colab environment to initiate cleaning and preprocessing. There are standard data cleaning and preprocessing processes to prepare the dataset to better perform in the machine learning model. The steps for data cleaning are as follows:

**Duplicate Records Deletion:** Duplicate records occur when similar records are present multiple times in the dataset due to data collection errors or human input mistakes. These duplicate records distort data analysis by providing certain data points with excessive weight. In the dataset, the number of duplicate records was 534, which has been dropped from the dataset.

**Handle Missing and Null Values:** After analyzing the ”CIC-MalMem-2022” dataset, there were no missing values or null values.

**Handle Unnecessary Columns:** Features that don’t contain meaningful information or have unnecessary values will introduce noise and increase computational complexity, which decreases the model’s performance. In the dataset, there were nine unnecessary columns with zero values that were dropped from the dataset.

**Outliers handling:** Outliers in a dataset can significantly impact statistical measures, and they also mislead predictions of machine learning models. In the dataset, there were some columns that had massive outliers. At first, feature-wise outliers were handled by using a capping method that had been applied to cap outliers by setting lower and upper thresholds. After analyzing the data distribution and histograms of all the features, the decision had been taken to handle class-wise outliers for every target class capping method that had been applied to cap outliers by setting lower and upper thresholds. For some extreme values, the Isolation Forest was applied to handle those extreme outliers. These two-step outlier handling methods gave a better data distribution among target classes.

### *C. Data Analysis*

The dataset contains three malware families, Ransomware, Spyware, Trojan and the benign class. The three of the malware families are further subdivided into five malware classes each, totaling fifteen different malware classes and the benign class.

Our dataset is a high-dimensional dataset consisting of 58,596 rows and 56 feature columns. In order to visualize it, we are required to plot it in a 2D or 3D space. Hence, we used the t-SNE (t-Distributed Stochastic Neighbor Embedding) visualization technique for both the family and individual classes.

t-Distributed Stochastic Neighbor Embedding is a non-linear dimensionality reduction technique used for visualizing high-dimensional data in 2D or 3D. It is capable of preserving local structures and minimizing KL divergence by converting pairwise similarities into probability distributions.

According to Figure 2, a clear distinction between binary class and malware classes can be seen, while there are some outliers in the case of both the binary and the malware families. This is also the reason why the binary classification of the dataset tends to be 100% accurate.

The t-SNE Visualization of the individual classes also shows a similar picture in Figure 3. Malware classes are generally mixed together, while benign classes are mostly separate.

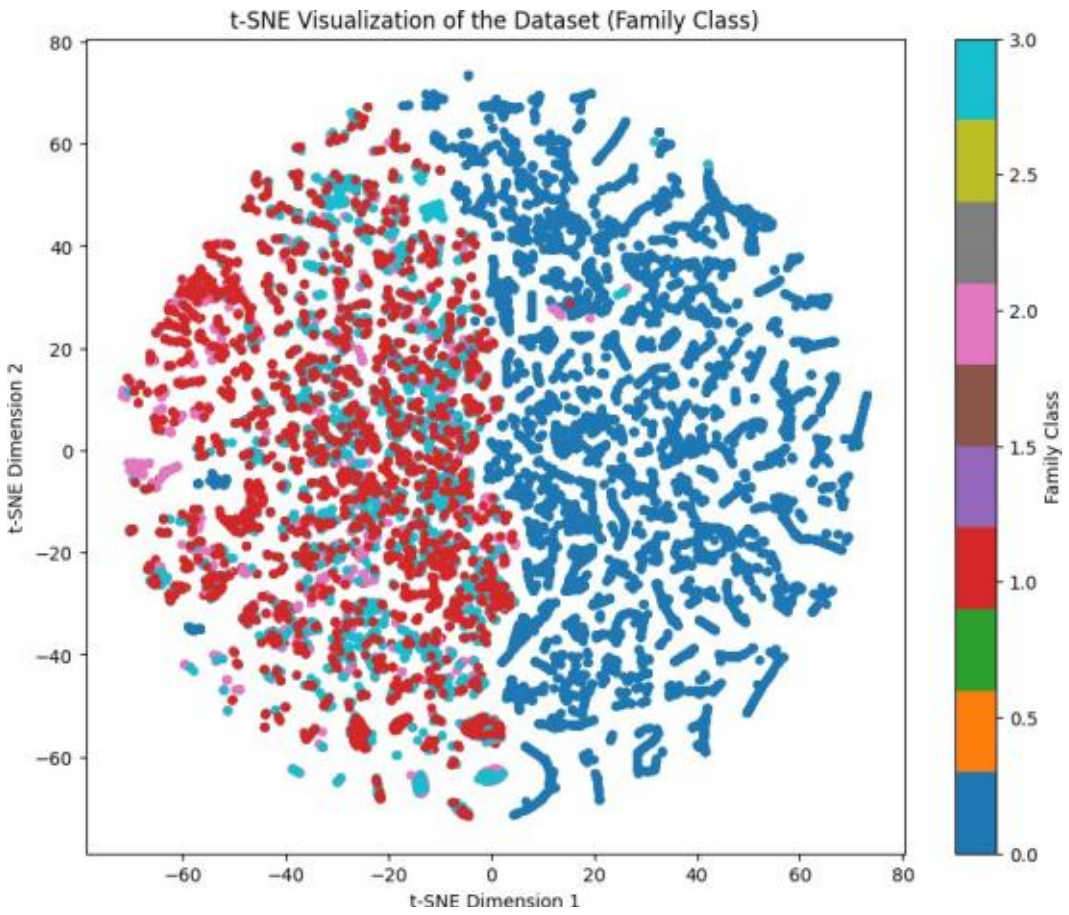

Fig. 2: t-SNE Visualization of the Family Class

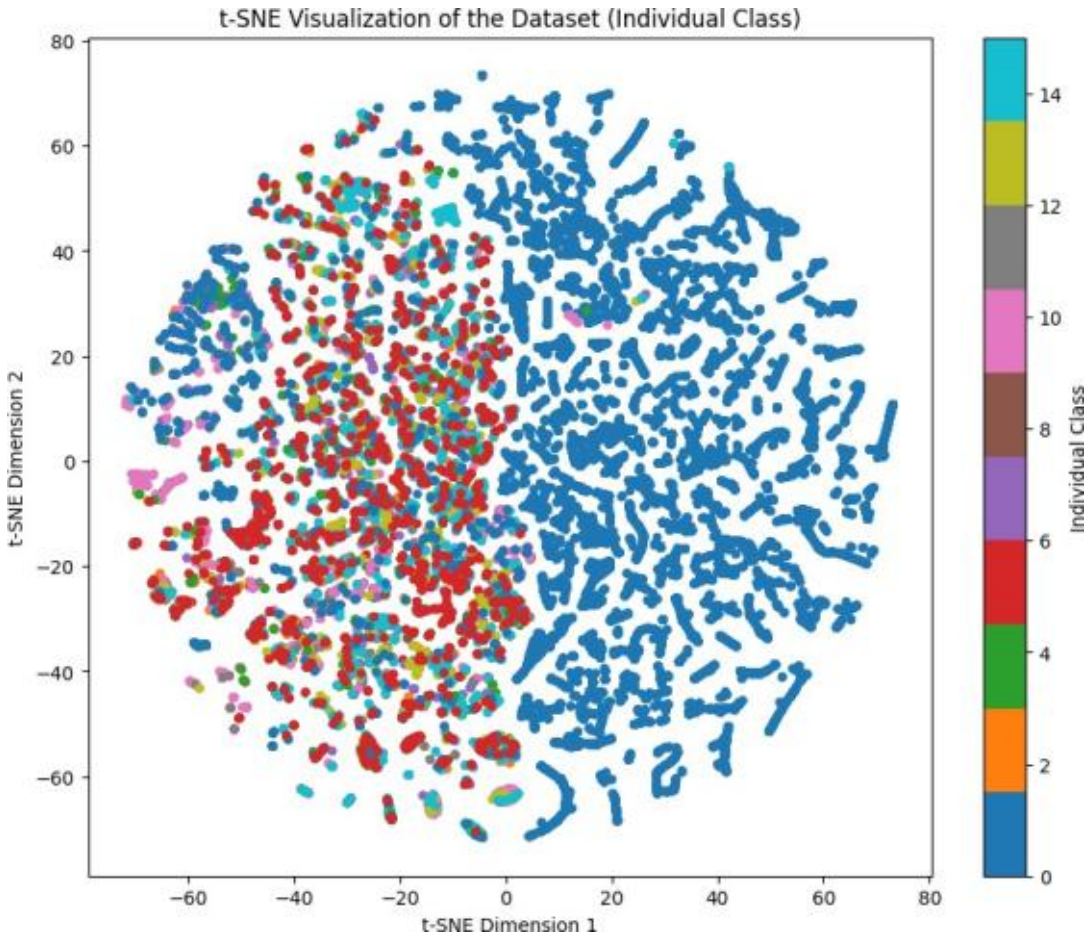

Fig. 3: t-SNE Visualization of the Individual Class

Although we can observe a higher percentage of overlap of the benign class with the malware classes. Reflecting a lower overall possible accuracy of the individual classes.

### *D. Data Balancing*

At the beginning of the data balancing process, the dataset had been divided into two parts, 80% for training and the remaining 20% for testing the dataset. We split the dataset before balancing to keep the testing set unique and untouched. To ensure a balanced dataset of all family class malware and individual class malware, a hybrid data balancing method was selected. To undersample the majority classes from the dataset, the Self-Organizing Map-based Undersampling (SOM-US) and, for the minority classes, interpolation between close samples in the feature space, the Synthetic Minority Oversampling Technique (SMOTE). This two-stage undersampling and oversampling hybrid technique effectively addresses the imbalanced class, which reduces the risk of overfitting and maintains the structure of the feature space.

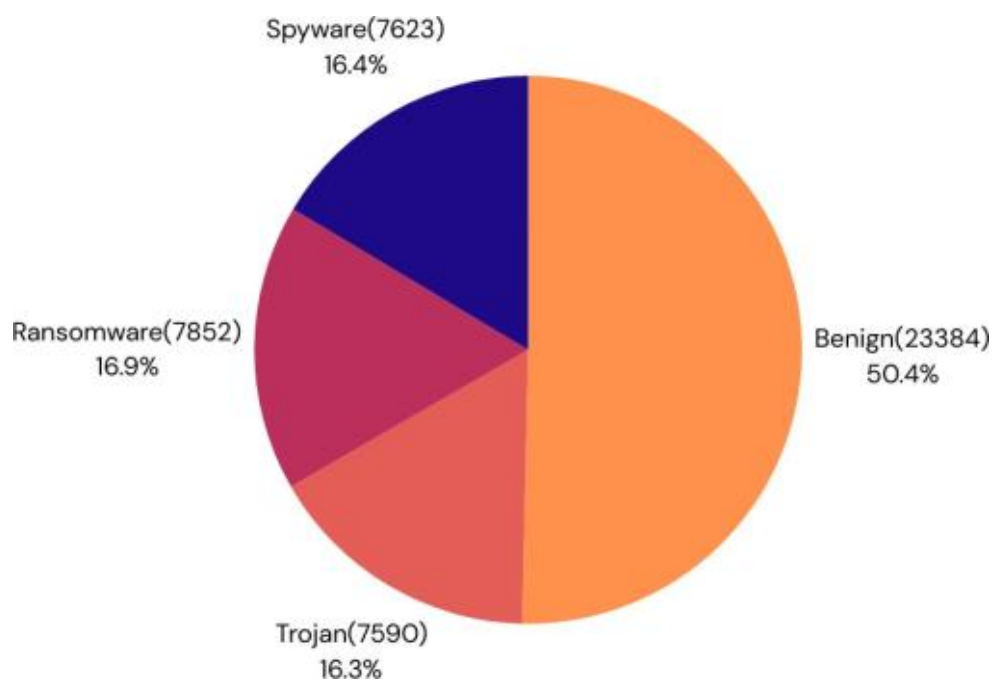

Fig. 4: Imbalanced Family Class

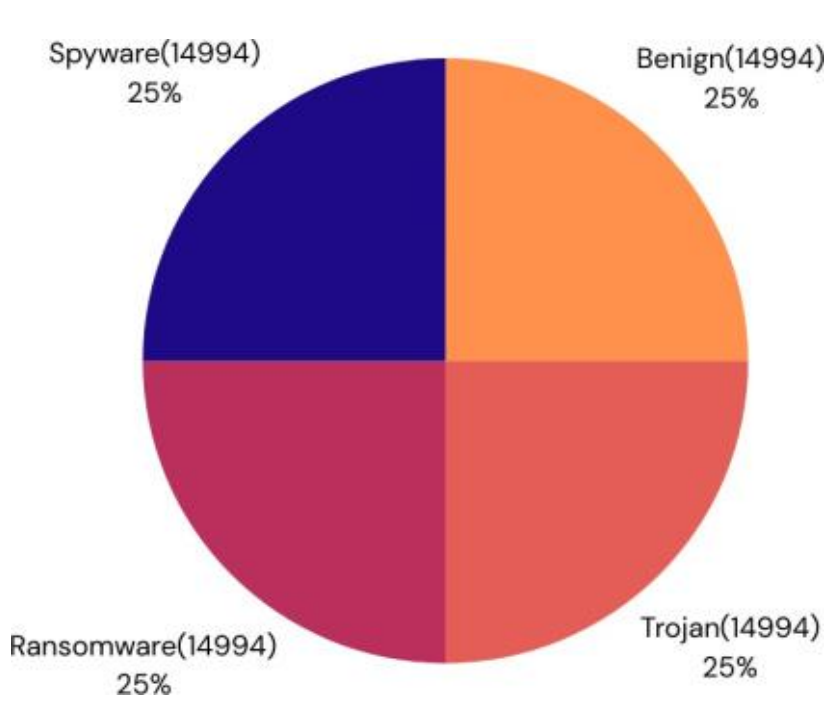

Fig. 5: Balanced Family Class

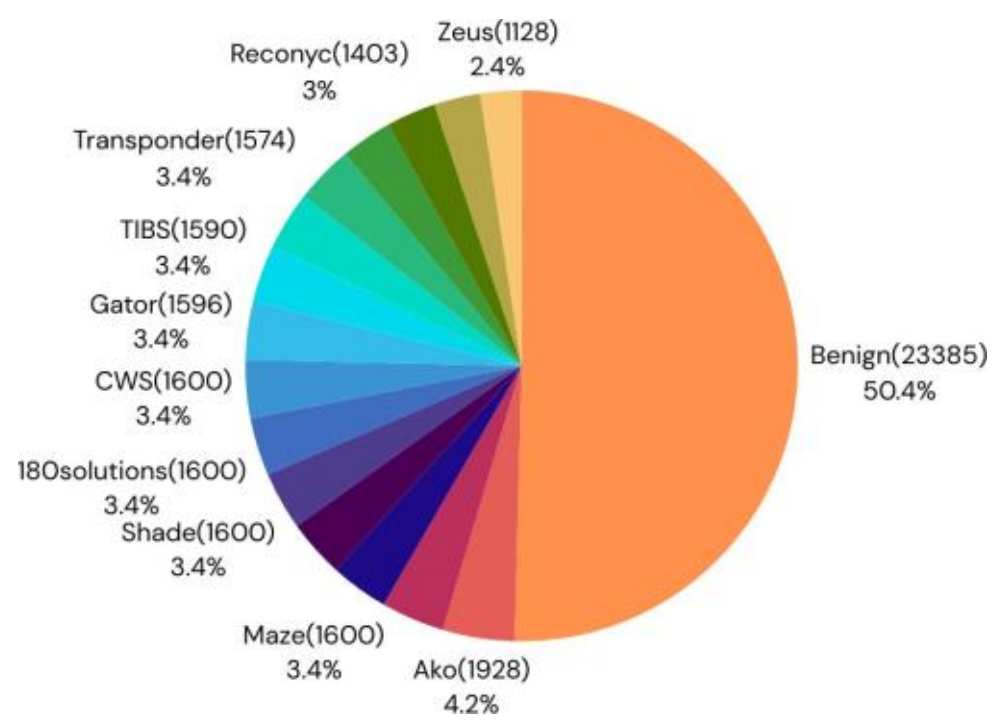

Fig. 6: Imbalanced Individual Class

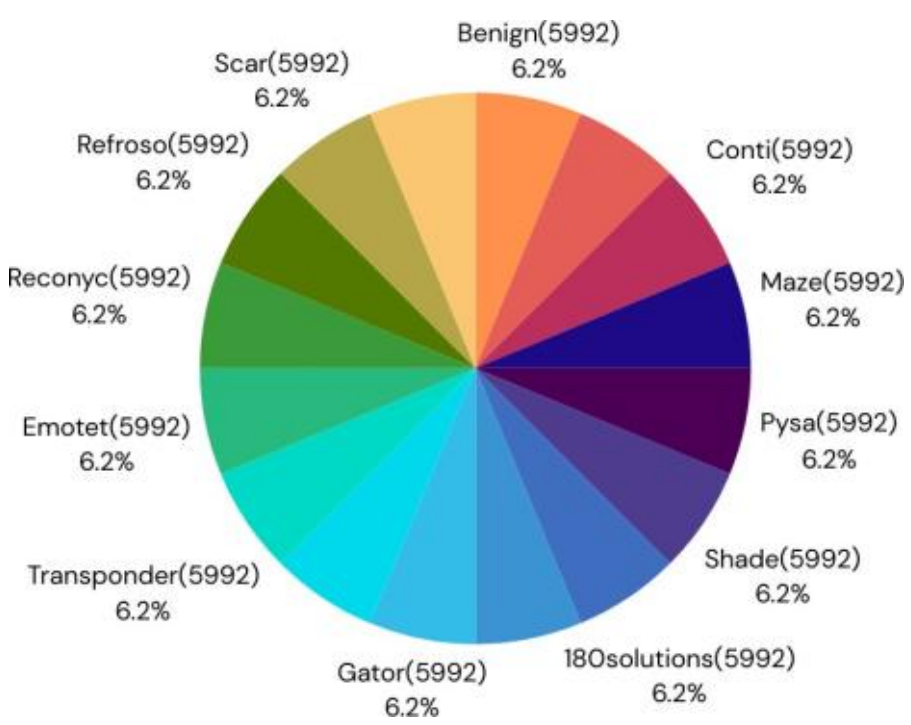

Fig. 7: Balanced Individual Class

**Self-Organizing Map with Undersampling (SOM-US):** SOM-US is a hybrid undersampling technique that uses Self-Organizing Maps and undersampling. It applies an unsupervised learning method of topological structure of the data that identifies and removes the redundant or informative samples, which gives better performance. From which representative samples are selected. This ensures the distribution of the majority class is retained.

**Synthetic Minority Over-sampling Technique (SMOTE):** SMOTE is an oversampling technique that uses k-nearest neighbors in order to interpolate between existing minority class instances. By using this method, SMOTE reduces overfitting possibilities as it introduces new synthetic data points. As we had observed, the benign class has a disproportionately high distribution in the dataset. Hence, simply running the classification models using the unbalanced dataset results in bias towards the benign class and skews the accuracy. For this, we incorporated data balancing methods, namely SOM-US (Self-Organizing Map with Undersampling) for undersampling the majority benign class and SMOTE (Synthetic Minority Over-sampling Technique) for oversampling the minority classes. These were implemented for both the classification of family and individual classes. The changes can be viewed in the figures 4 to 7.

### *E. Feature Selection*

The next step of data preprocessing, which is feature selection, was conducted. In order to improve the model's performance and efficiency, we used a feature reduction technique, PCA, and two feature selection techniques, the firefly algorithm and the genetic algorithm.

**Principal Component Analysis (PCA):** Principal Component Analysis (PCA) is a commonly used technique for reducing dimensionality of a dataset. It transforms a large dimensional dataset into a lower dimension while trying to preserve as much variance as possible. PCA overall helps in reducing the computational complexity by removing redundancy and increasing the model's performance by removing the correlated features. In our study PCA was applied to the "CIC-MalMem-2022" dataset after cleaning, which included the removal of outliers. Since PCA is sensitive to outliers, handling extreme values ensures that the principal components show meaningful variance in the data. After performing PCA, 21 principal components were obtained, representing the most significant directions of variance in our dataset. PCA-based feature selection was unable to enhance the performance of the model even if it reduced dimensions of the dataset. Alternative feature selection methods were needed because the principal components were able to capture variance but were not relevant for our model training.

**Firefly Optimization Algorithm (FOA):** The Firefly Op-

| Index | Feature | Description |
|---|---|---|
| 3 | pslist.avg_threads | Average usual number of threads per process |
| 6 | dlllist.ndlls | Number of all loaded DLLs libraries |
| 7 | dlllist.avg_dlls_per_proc | Average amount of all loaded DLLs libraries for each process |
| 9 | handles.avg_handles_per_proc | Average amount of handles per process |
| 11 | handles.nfile | Indicate the numbers of opened and created resources |
| 14 | handles.nkey | Number of the opened registry keys |
| 17 | handles.nsemaphore | Number of the shared resources controls |
| 18 | handles.ntimer | Number of timers handles (e.g., periodic check and time scheduling) |
| 19 | handles.nsection | Number of sections that are shared with other processes |
| 20 | handles.nmutant | Number of mutex in system |
| 21 | ldrmodules.not_in_load | Number of mismatched loaded module order |
| 22 | ldrmodules.not_in_init | Number of not initialized modules |
| 23 | ldrmodules.not_in_mem | Number of modules that do not reside in process memory |
| 25 | ldrmodules.not_in_init_avg | Average of unmatched initialized modules |
| 28 | malfind.commitCharge | Total virtual memory used by all processes (Commit Charge) |
| 29 | malfind.protection | Number of protections |
| 30 | malfind.uniqueInjections | Number of unique injections |
| 38 | psxview.not_in_pslist_false_avg | Average false ratio of the process list |
| 46 | svcscan.nservices | Number of all services |
| 47 | svcscan.kernel_drivers | Number of all drivers in the kernel |
| 52 | svcscan.nactive | Number of all service processes that are actively running |
| 53 | callbacks.ncallbacks | Number of all callbacks |

Fig. 8: Selected Features List

timization Algorithm (FOA) is a metaheuristic optimization technique that is inspired by the natural flashing behavior of fireflies. It is mainly used for feature selection and solving the high-dimensional dataset problems because it mimics how fireflies are drawn to brighter individuals, directing the search toward optimal solutions. The firefly optimization algorithm is used where traditional methods are ineffective due to the large search size of the data space. It mainly helps to identify the features that are most relevant in large dimensional datasets and reduce model complexity while preserving all the important information. In the study, FOA was implemented for feature selection. The algorithm selected all important features, aiming to improve the model efficiency by reducing the dimensions of our large dataset. However, after running different machine learning models, the selected features did not give the desired improvement in performance. Thus, other feature selection methods were explored in the study.

**Genetic Algorithm (GA):** The genetic algorithm is a bioinspired optimization technique that is similar to the process of natural selection. It basically evolves a population of candidate solutions over multiple generations, selecting only the best ones based on a defined fitness function. This makes the Genetic Algorithm (GA) a very effective feature selection method in a high-dimensional dataset, where traditional search methods are ineffective due to large search size. The Genetic Algorithm (GA) helps to identify the most useful features by simulating the evolutionary process, such as selection, crossover, and mutation. It helps improve model performance by reducing redundancy while maintaining essential information. The Genetic Algorithm (GA) balances between searching diverse feature combinations and refining the best solutions to find an optimal subset of features to improve a model's performance. In this study, the Genetic Algorithm (GA) was made to run for 10 generations using the weighted F1 score as the fitness function to evaluate the feature subset. After running the machine learning models, the selected features did give the desired output from the feature selection process by improving the model's performance. This indicated that the Genetic Algorithm (GA) for feature selection was more effective than other methods used in enhancing model performance and reducing the complexity of the dataset. In conclusion, for this study, the genetic algorithm performed the best for feature selection over PCA and Firefly. The genetic algorithm selected 22 features out of a total of 46 to be relevant and important to improve the model's performance. We obtained the feature descriptions from Aljabri et. al. [19] (2024). The selected features are given in the Figure 8.

### F. Resource Feasibility of Feature Extraction

In order to successfully implement the model in resource-constrained environments, we not only require the size of the model to be optimized, but we also require the feature extraction process triggered by an event or executable that is incoming to work under limited computational environments. Therefore, a consideration of the real-time viability of extracting the chosen features from Figure 8 is necessary. Features such as ”pslist.avg_threads”, ”dlllist.ndlls” etc., are obtainable through lightweight Linux command-line tools such as the /proc filesystem interface as documented by The Linux Kernel Organization [20]. On the other hand, the features in ”ldrmodules” group (e.g., ”ldrmodules.not_in_load”, ”ldrmodules.not_in_init”, ”ldrmodules.not_in_mem”, ”ldrmodules.not_in_init_avg”) and ”psxview.not_in_pslist_false_avg” and ”callbacks.ncallbacks”, were not feasible to extract for IoT devices with less than 1 GB RAM. This can be seen in the implementation described by Mosli et. al. [21] (2017), where utilities such as Cuckoo Sandbox, INetSim, Volatility Framework, and VirusTotal required a minimum of 1 GB of free memory for effective execution of the feature extraction process. Therefore, we came to the conclusion that 6 among the 22 selected features were not extractable in IoT devices with less than 1 GB of RAM.

Although our model itself is capable of running in environments offering as little as 2.5MB of memory (as later discussed in Chapter 5), the existing tools for feature extraction require considerably more than that. For us, this creates a limitation for real-world deployment in resource-constrained environments.

### G. Summary of Preprocessed Data

In the study, the “CIC-MalMem-2022” dataset was used, which is crucial not only because of its size but also for its focus on the dynamic aspects of malware behavior in system memory. For the preprocessing and the development of a model, the dataset has been loaded in the Google Colab environment to initiate cleaning and preprocessing. The first section of data preprocessing is the data cleaning part. There were 534 duplicate records found and thus removed from the dataset. After that, missing values and null values were checked. Then some unnecessary columns were found and was dropped from the dataframe. Outlier handling is the most important part of data cleaning. To handle outliers in the dataset, class-wise outliers were checked and capped by setting lower and upper thresholds. For some extreme values, the Isolation Forest was applied with 0.1% contamination, which is quite a small amount to handle outliers. Secondly, the data points for different classes of data distributions were analyzed by using t-SNE visualization and correlations by using heatmaps. Thirdly, the category columns have been divided into three target columns, such as Binary Class, Family Class, and Individual Class, where Binary Class was for binary classification and Family and Individual Classes were for multiclass classifications. After that, in the dataset, label encoding was applied for machine learning models. At last, the MinMaxScaler was applied to normalize both training and testing feature sets, and as the dataset is not balanced, a decision was taken to balance the dataset by using SOM-US for undersampling (15,000 for malware family classifications and 6,000 for individual malware classifications) and SMOTE for oversampling the minority classes.

## IV. Model Design and Methodology

### A. Methodology Overview

In this research paper, a machine learning approach had been adopted to detect obfuscated malware for resource-constrained environments mostly targeting IoT. The approach followed data preprocessing, feature reduction, and comparative analysis of various classification models to develop the most effective and efficient machine learning model. The methodology was followed to ensure the high accuracy of detection, reducing false positives and optimal computational efficiency to develop a lightweight model. The overall process of the proposed model can be observed in the figure 9.

The overall process comprises multiple steps, such as problem definition, dataset collection, data preprocessing, feature reduction, data analysis, model experimentation, evaluation, and final model selection. These steps are step-by-step and refined based on the performance and computational constraints.

The malware detection was initially formulated as a supervised multi-classification task where the main objective was to predict the malware families of 3 different classes and 15 multiple individual classes based on the extracted behavioral and memory-based features. The "Obfuscated-MalMem2022" dataset has been used that contains memory dump features as well as malicious samples. To ensure the authenticity of the classification models, included data cleaning, label encoding, normalization, and class balancing for both family class classifications and individual class classifications. To balance the dataset, Synthetic Minority Oversampling Technique (SMOTE) and Self-Organizing Map-based Undersampling (SOM-US) based undersampling were applied. To experiment, multiple machine learning algorithms were selected, such as, Random Forest (RF) classification, which is an ensemble tree-based classification model and is known for its robustness; LightGBM classification, which is a gradient boosting algorithm that optimizes speed and memory usage; Decision Tree (DT), which is a baseline interpretable model; and K-Nearest Neighbor (KNN), which is a distance-based algorithm.

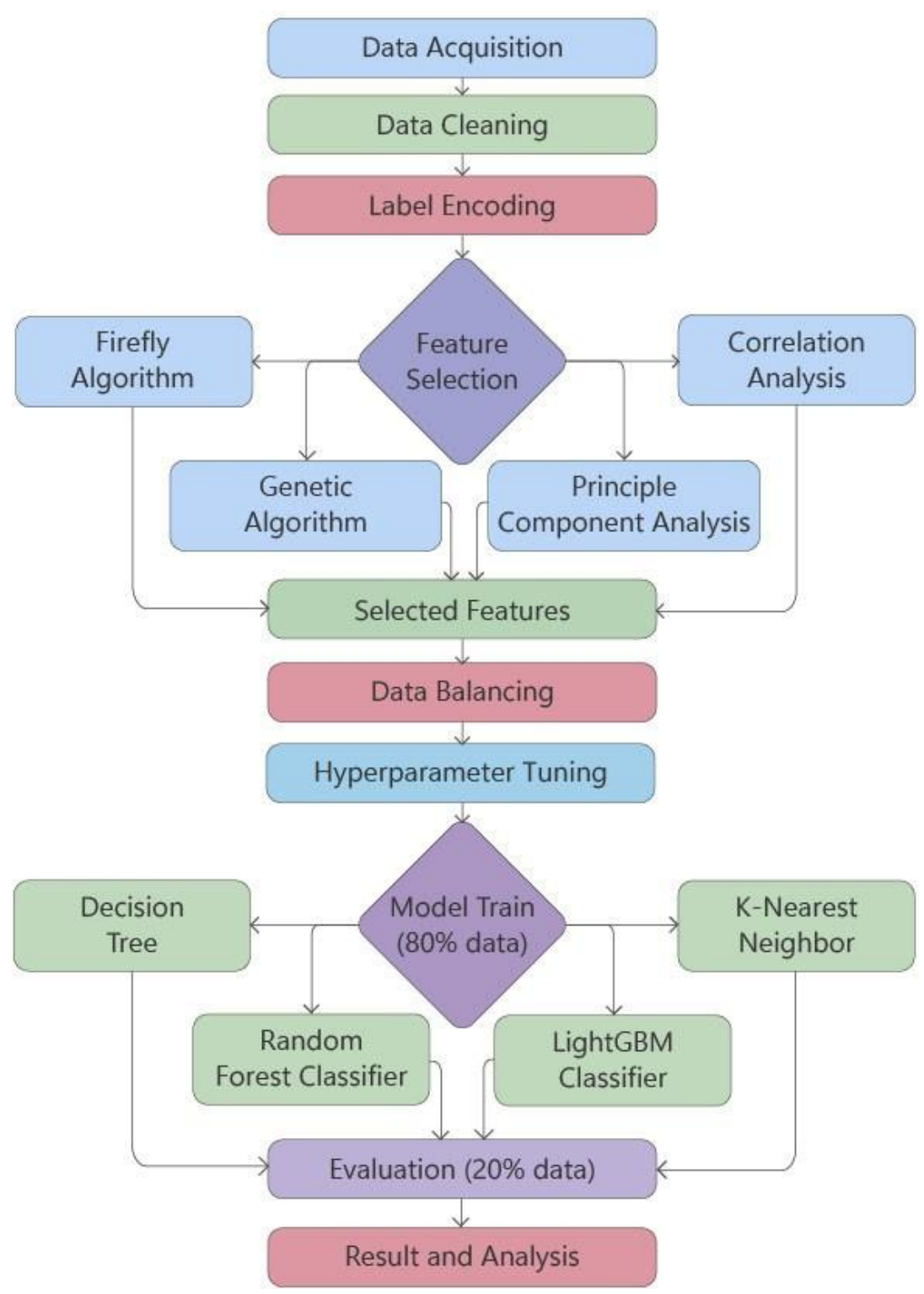

Fig. 9: Overall Workflow

All the models were trained using 80% training and 20% testing splits to maintain the consistency in the evaluation. Performance metrics, for example, accuracy, precision, recall, and F1-score, as well as the confusion matrix, had been used for evaluating the models. Evaluating the models, the Random Forest (RF) model consistently outperformed other models in the multiclass classification, where the accuracy, precision, and F1-score were higher.

Additionally, to consider the performance of models, the Random Forest (RF) model with hyperparameter tuning performed better than other models because of its computational efficiency. Although LightGBM also demonstrated competitive results, it is sensitive to parameter tuning, which is less stable.

### B. Model Architecture

This section describes the overall outline strategy, preprocessing pipeline, and the selected model architectures to detect multi-class malware detection for both family class and individual class. The main target of this research is to build an efficient and lightweight model to achieve higher classification accuracy, which takes less training time and limited memory usage.

The models were selected based on the performance of the accuracy and the computational efficiency. Four machine learning classification models were selected: Random Forest (RF), Light Gradient Boosting Machine (LightGBM), Deci-sion Tree (DT), and K-Nearest Neighbor (KNN). All the models are computationally light; because of that, it takes less time to train, which is sustainable for resource-constrained environments. Among these four models, Random Forest (RF) and LightGBM were selected because of their ensemble learning capabilities and robust performance, while DT and KNN were selected for their simpler baselines. The comparative use of these models identifies the most efficient and accurate classifier for real-world deployment.

**Random Forest:** As illustrated in the Figure 10 by Chen et al. [22], the training set gets split into multiple subsets through bootstrap sampling. Then each subset is fed to independent decision trees. The trees process the dataset and learn predictions. When the test set is pushed for classification, each tree predicts, and through majority voting, the classification is finalized. Additionally, a comprehensive evaluation phase following model training was implemented, during which k-fold cross-validation was used to gauge the model's performance across various dataset subsets to ensure that the model performs consistently well. The model's effectiveness and dependability are evaluated using performance indicators like accuracy, precision, recall, and F1-score. Later, the model is trained with a focus on tuning its hyperparameters, such as the number of trees, the maximum depth of the trees, and the minimum number of samples required to split a node. These parameters are crucial for the model to effectively learn from the complex patterns without overfitting or underfitting. The Random Forest is particularly suitable for this task due to its ensemble nature, where multiple decision trees vote on the final output, improving the model's accuracy and reliability. Moreover, as it is a machine learning language, it is significantly lightweight and requires low computational resources.

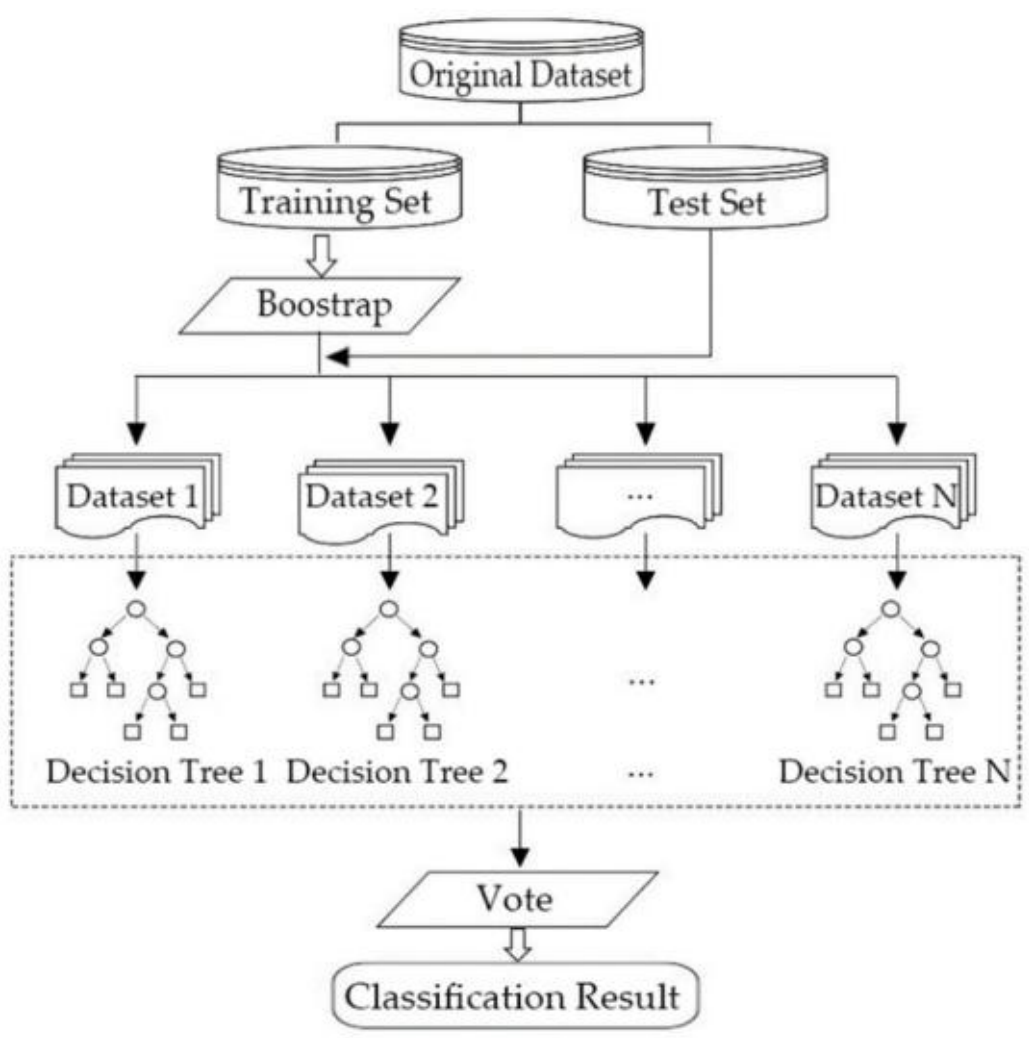

Fig. 10: Random Forest

**Light Gradient Boosting Machine:** Light Gradient Boosting Machine (LightGBM) is a quick and efficient gradient boosting algorithm intended for large-scale databases. Unlike standard models, it grows trees leaf by leaf, as seen in Figure 11, selecting the branch with the greatest loss reduction, which improves accuracy but may lead to overfitting. To speed up training, it employs Gradient-based One-Side Sampling (GOSS), which retains important samples while rejecting the less valuable ones, and Exclusive Feature Bundling (EFB), which helps reduce feature dimensionality by grouping non-overlapping features. LightGBM does not check every possible split; instead, it builds histograms of the feature values, which reduces the computation time. LightGBM also supports parallel processing and CPU acceleration, which make it useful for large-scale datasets. Additionally, LightGBM also has early stopping and regularization, which helps to prevent overfitting. These optimizations make LightGBM much faster and more memory-efficient than traditional gradient-boosting models. The LightGBM framework is used to train the model, with an emphasis on adjusting hyperparameters such as the maximum depth, learning rate, and number of leaves. The model's performance is assessed using k-fold cross-validation and metrics like accuracy, F1-score, precision, and recall. Following model tweaking and optimization, the model is evaluated on the independent test set to determine its reliability. Just like Random Forest, LightGBM is also a machine learning language that is lightweight and efficient.

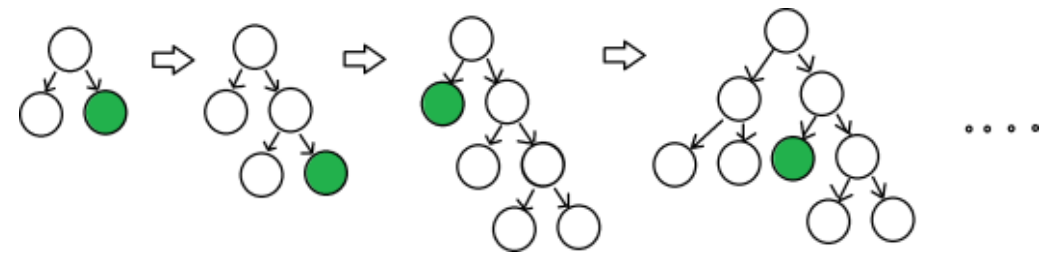

Fig. 11: LightGBM

**Decision Tree:** Decision Tree is a hierarchical, flowchart-like structure. It creates a tree where nodes represent a decision based on the feature values that result in the best purity gain. These decisions result in splitting of the dataset and, therefore, honing in on the prediction class.

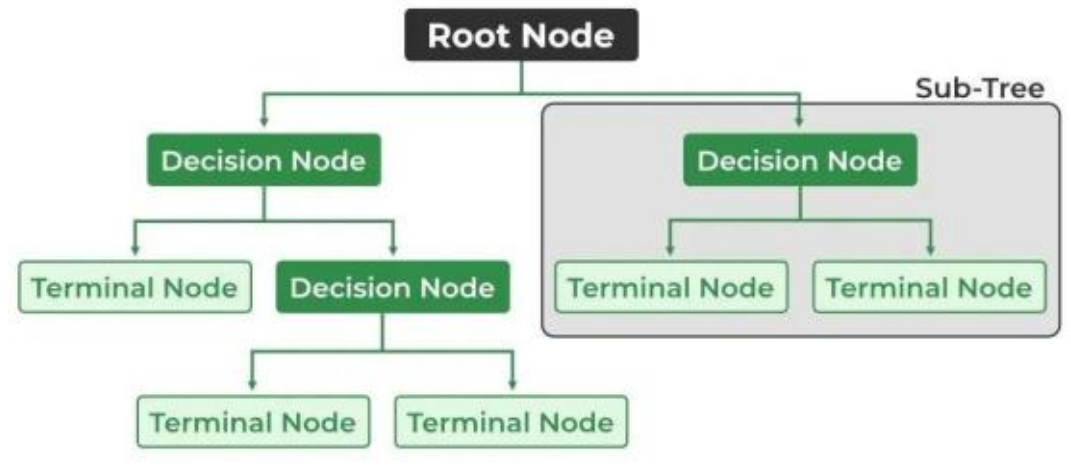

Fig. 12: Decision Tree

When the remaining samples belong to a singular target class or maximum depth is reached, the tree completes its creation. During the test phase, the samples traverse the tree, picking the next node based on the values till they reach the terminal node, as seen in Figure 12 and therefore, decide on a target class. The previously discussed Random Forest is essentially a model that combines the predictions of multiple decision trees to reach a final decision. Decision Tree is a light and valuable architecture for our objective.

**K-Nearest Neighbour:** KNN is a non-parametric classifier. It uses the proximity of the data points to make predictions during the testing phase. Meaning, during training, it does not build a model. It uses distance metrics such as Euclidean, Manhattan, Cosine, etc. to calculate the distance between the test sample's feature values and its neighboring training samples. The number of samples considered to determine the prediction relies on the K-value. Through a majority voting mechanism, it then classifies the test sample based on its neighbor's classes. As per figure 13, we can see Class 2 would be selected as the predicted class because it had the majority. Due to its unique method of classification as compared to the other tree-based, model-driven ones, implementing KNN allows us to judge the inherent structure of the dataset in terms of classification without heavy model assumptions.

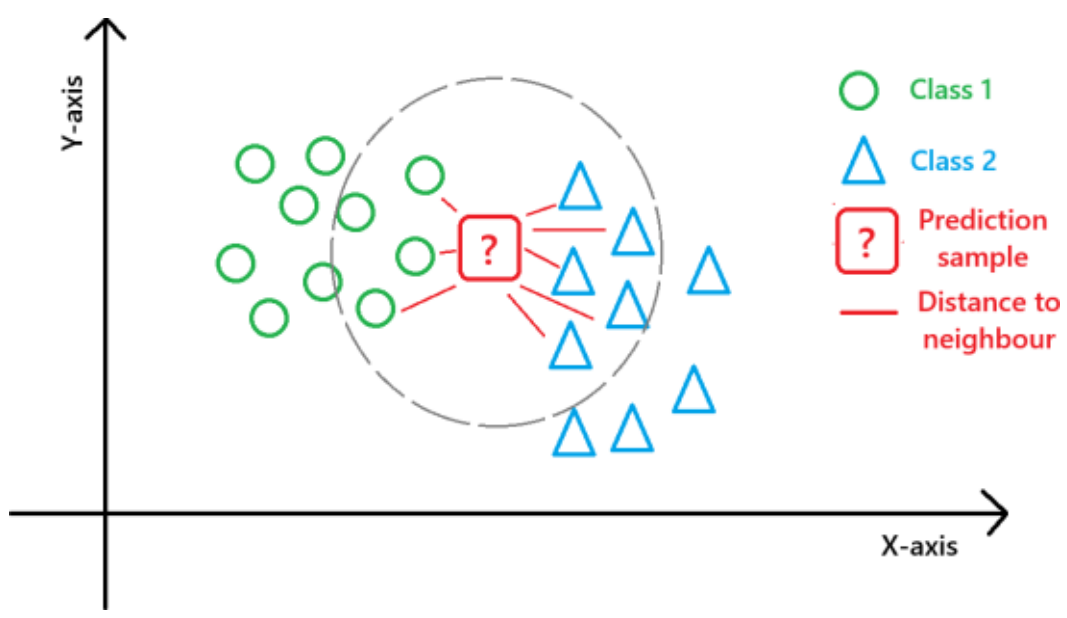

Fig. 13: K-Nearest Neighbour

### C. Implementation of Selected Design

**Environment Setup:** Both family class classification and individual class classification were developed using Python libraries such as scikit-learn, pandas, numpy, matplotlib, seaborn, minisom, imblearn, and LightGBM. Google Colab was used as our development environment. The CSV file of the dataset was uploaded to Google Colab and then transformed into a pandas dataframe. Then we implemented the aforementioned data preprocessing steps in Chapter 3. Thus, the data set was ready for the classification model.

**Training Phase:** In the training phase, each of the classification models was trained with the preprocessed 80% data from the dataset using the machine learning models. In the training phase, metrics such as accuracy, loss curves, and class-wise performance were deeply analyzed to avoid overfitting.

**Testing Phase:** In the test phase, 20% split data from the data set before balancing was selected and evaluated using the trained model to assess performance consistency. Performance metrics, including accuracy, precision, recall, and F1 score, have been used for each classification, and for the false positive result, confusion matrices have been used. To evaluate the classifier's ability, the ROC-AUC scores were used.

**Parameter Tuning and Configuration:** To achieve optimal performance, each of the models was fine-tuned using the Grid Search algorithms, manual testing, Optuna and cross-validation techniques. Most significant parameters, such as number of trees "n_estimators", tree depth "max_depth", and minimum split size "min_samples_split" were deeply varied for optimal performance in Random Forest (RF), LightGRM, and Decision Tree (DT) models. For the K-Nearest Neighbour (KNN) model, different values of "k" and distance metrics were tested to get the most optimal performance. The hyperparameter tuning focused not only on maximizing the accuracy, precision, and F1-score but also on minimizing the training time and minimum memory usage to make the model lighter for resource-constrained environments.

Overall, the implementation is evaluated with the trained as well as validated lightweight models, which are capable of detecting obfuscated malware with better accuracy as well as fewer false positive rates. The hybrid data balancing techniques using SMOTE and SOM-US, the comparative analysis of multiple classification models gave a real-world development in a resource-constrained environment.

## V. Result Analysis

Four classification models were used to predict the classes both on a malware family level and on an individual malware level. Then, cross-validation was implemented to maintain consistency among the outcomes. Furthermore, hyperparameter tuning through Grid Search, Optuna and manual testing was done to reach the optimal parameters for each of the classifiers. The classification findings for malware family class and individual malware class will be discussed in the following sections, respectively.

### A. Malware Family Classification

From the classification reports in Figure 14 and Figure 16 as well as the confusion matrices in Figure 15 and Figure 17, it can be observed that the common misclassifications are between Ransomware and Trojan (class 1 and class 3), particularly when the real malware class is Trojan but the models predict it as ransomware. Decision Tree, in particular, has the highest misclassification in this case. Meanwhile, LightGBM struggled to correctly classify spyware. In general, KNN had the overall worst misclassification rate. From this, it can be understood that the common misclassification is between Ransomware and Trojan. Decision Tree struggles in classifying Trojan correctly, and LightGBM struggles in classifying Spyware correctly. All the models are able to correctly predict the benign class, meaning they have a 100% accuracy in terms of binary classification.

As illustrated in Table I, Random Forest outperformed the other classifiers by at least 2%. It reflects its effectiveness due to its ensemble characteristics. But there is also a downside to the RF model that we observed. The average training time as well as the average testing time for 80% of the dataset and 20% of the dataset, respectively, were calculated. The overall model size was also calculated.

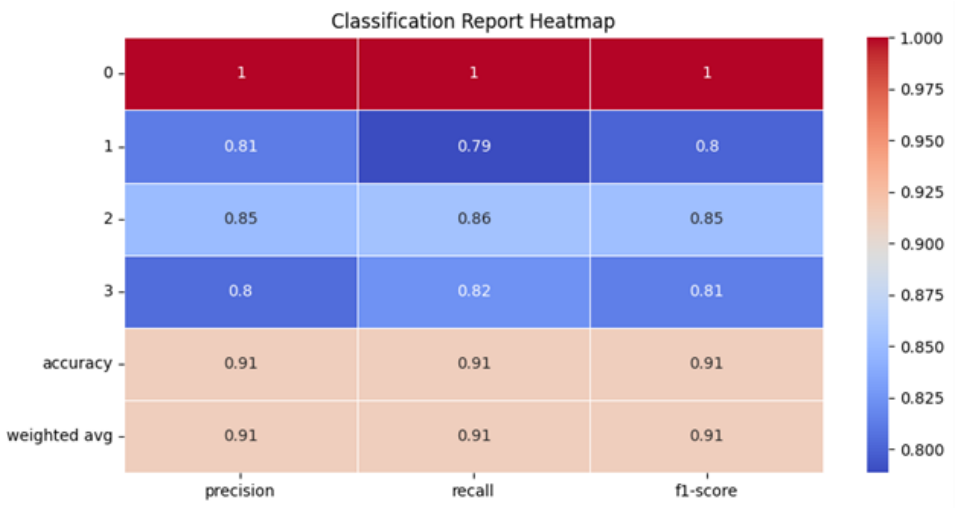

Fig. 14: Random Forest Classification Report (Malware Family)

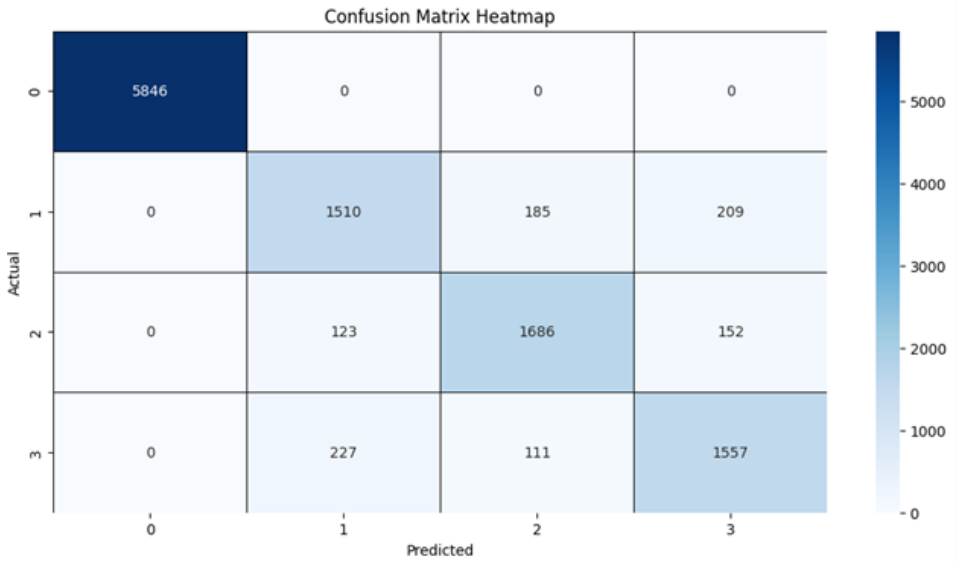

Fig. 15: Random Forest Confusion Matrix (Malware Family)

From Table II, it was noticed that because of Random Forest's ensemble characteristic, it requires significantly more training time and storage space. While training time in practical scenarios does not bear as much significance as testing time does, model size is another case. As the focus is on IoT devices, the constraints of them must be considered. While some high-end IoT devices such as Raspberry Pi 4 and NVIDIA Jetson Nano can store the models locally, low-end

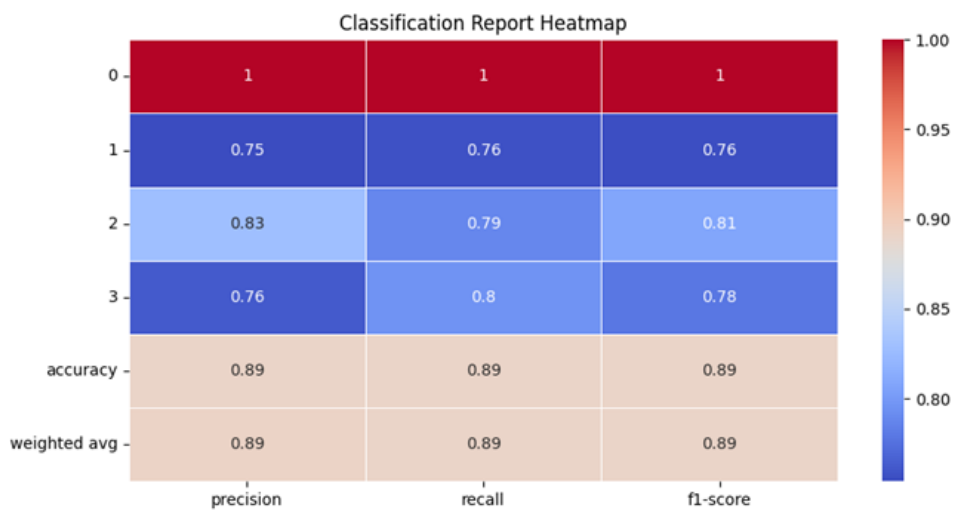

Fig. 16: LightGBM Classification Report (Malware Family)

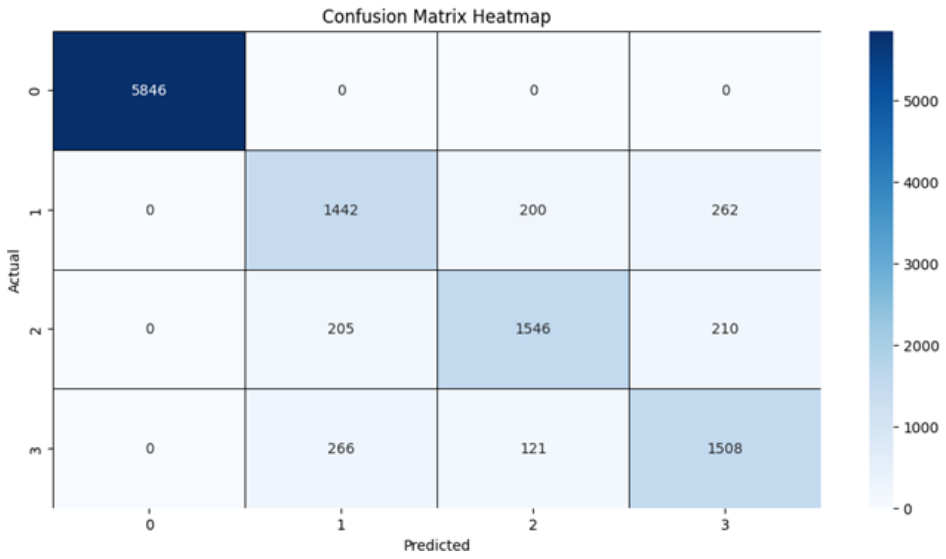

Fig. 17: LightGBM Confusion Matrix (Malware Family)

| Classification Model | Precision | Recall | F1-score | Accuracy |
|---|---|---|---|---|
| Random Forest | 0.911 | 0.912 | 0.912 | 0.912 |
| LightGBM | 0.891 | 0.891 | 0.891 | 0.891 |
| Decision Tree | 0.888 | 0.888 | 0.888 | 0.888 |
| K-nearest Neighbors | 0.852 | 0.852 | 0.852 | 0.852 |

TABLE I: Classification model performance metrics (Malware Family)

IoT devices such as ESP32 or STM32F4, which can store 4 MB and 2 MB, respectively, cannot store the Random Forest model and K-nearest Neighbors model locally. Therefore, from the findings, some conclusions were made.

| Classification Model | Training Time | Testing Time | Model Size |
|---|---|---|---|
| Random Forest | 8.2 s | 0.07 s | 34.48 MB |
| LightGBM | 2.075 s | 0.078 s | 0.5 MB |
| Decision Tree | 1.18 s | 0.003 s | 0.5 MB |
| K-nearest Neighbors | 0.01 s | 5.94 s | 10.53 MB |

TABLE II: Resource usage of classification models (Malware Family)

K-nearest Neighbors not only had a significantly larger model size, but it also had very low accuracy. Moreover, due to its nature of handling train and test data, it does most of its work in the training portion of the process. Hence, having a significant delay in real-life scenario classification creates a major vulnerability window. Therefore, KNN has been concluded to be unsuitable. Random Forest can be used in high-end IoT devices locally while being utilized through cloud-based detection or edge gateway detection. Where the metadata of the incoming event is sent to the cloud or a central gateway that will run the classification result there. Then relay the necessary steps the IoT devices must take depending on the predicted result. Meanwhile, Decision Tree and LightGBM produced similar results. LightGBM has a slightly higher accuracy, while Decision Tree has a faster prediction time. Each of the models demonstrated their strengths and constraints. Therefore, depending on the specifications of the IoT devices and the priority of the user, the appropriate model can be chosen.

### B. Individual Malware Classification

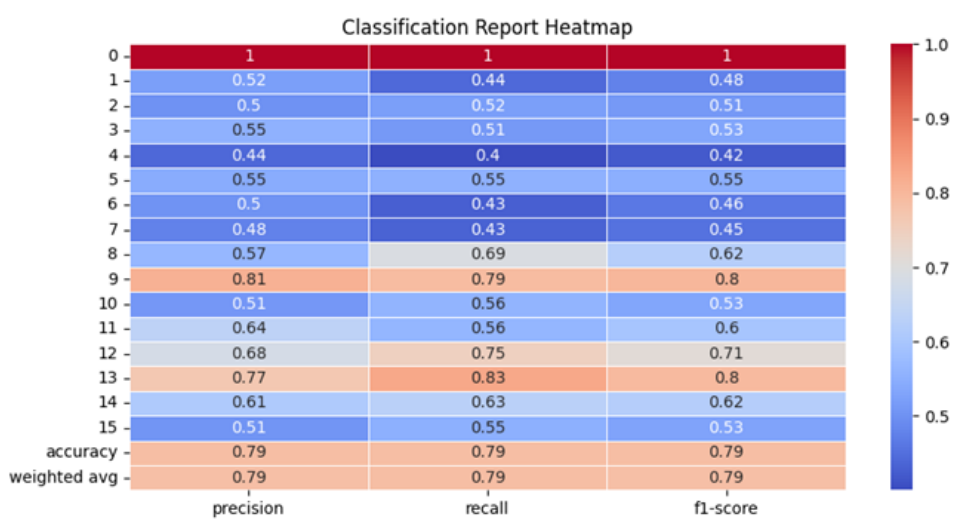

Fig. 18: Random Forest Classification Report (Individual Malware)

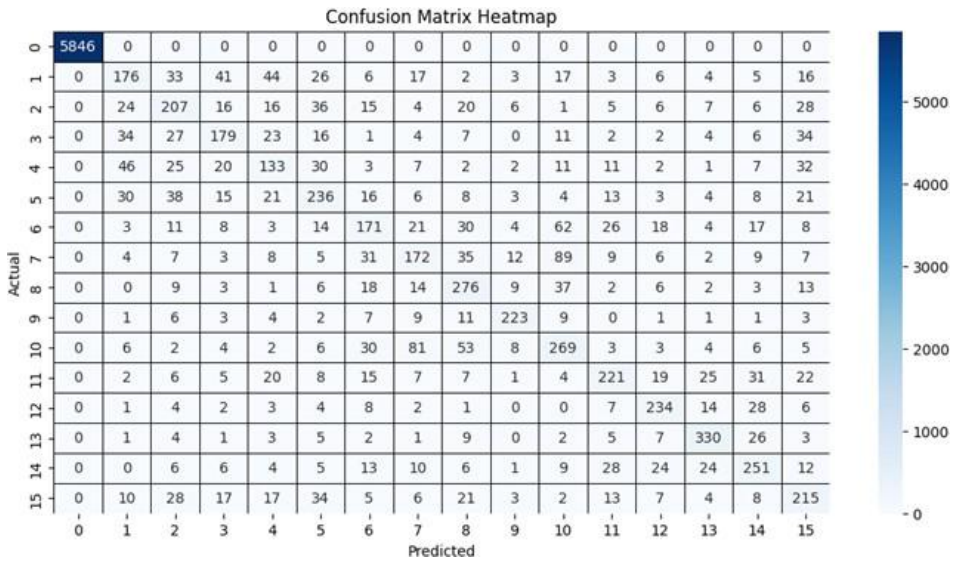

Fig. 19: Random Forest Confusion Matrix (Individual Malware)

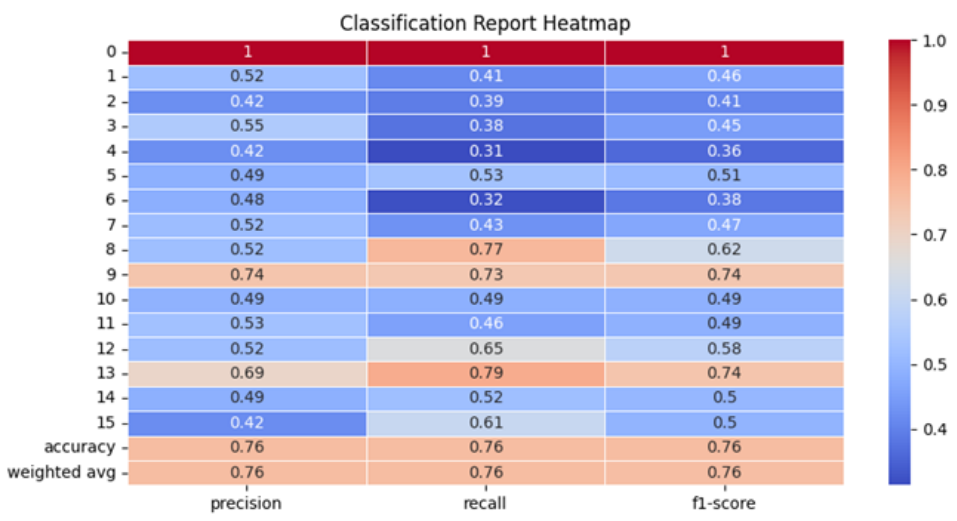

Fig. 20: LightGBM Classification Report (Individual Malware)

From the classification reports in Figure 18 and Figure 20 and the confusion matrices in Figure 19 and Figure 21, it can be observed that some classes are classified correctly more so than others, such as classes 9, 12, 13, and such. This consistency exists across the classification models. The lowest-performing individual malwares belonged to the Ransomware family, implying malwares that fall under Ransomware are less differentiable than those of Spyware and Trojan.

As seen in Table III, similar results to malware family classification were observed, where random forest produced the best result, followed by LightGBM, Decision Tree and K-nearest Neighbors respectively.

The model size of Random Forest makes it unsuitable for local implementation for IoT devices. Therefore, it must be utilized through a gateway or cloud-based detection. Unlike in the case of malware family classification, LightGBM is clearly the better model between it and Decision Tree. As we see in Table IV, it has a smaller model size and shorter testing time while producing higher accuracy than Decision

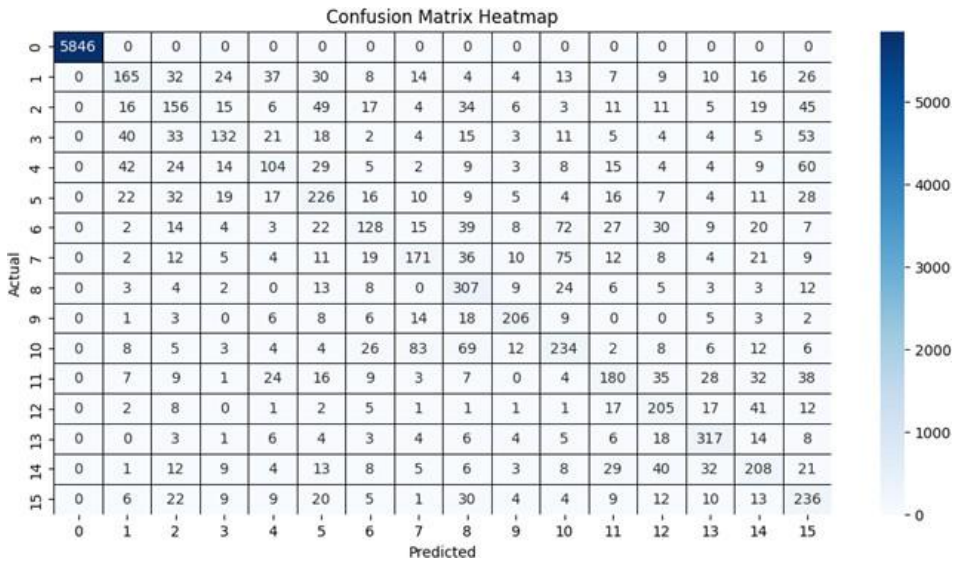

Fig. 21: LightGBM Confusion Matrix (Individual Malware)

| Classification Model | Precision | Recall | F1-score | Accuracy |
|---|---|---|---|---|
| Random Forest | 0.788 | 0.787 | 0.787 | 0.787 |
| LightGBM | 0.761 | 0.760 | 0.760 | 0.760 |
| Decision Tree | 0.744 | 0.743 | 0.743 | 0.743 |
| K-nearest Neighbors | 0.710 | 0.709 | 0.709 | 0.710 |

TABLE III: Classification model performance metrics (Individual Malware)

Tree. KNN fails due to the same reasons as the malware family classification. Taking more time during tests and generally having worse results than the other models. Therefore, for individual classes, it has been concluded that the Random Forest and LightGBM classifiers are suitable for the desired outcome. The use of one over the other depends on the specifications of the IoT devices and the priority of the user.

| Classification Model | Training Time | Test Time | Model Size |
|---|---|---|---|
| Random Forest | 24.9 s | 0.201 s | 234.71 MB |
| LightGBM | 12.65 s | 0.302 s | 2.11 MB |
| Decision Tree | 3.45 s | 0.005 s | 3.42 MB |
| K-nearest Neighbors | 0.01 s | 6.35 s | 16.83 MB |

TABLE IV: Resource usage of classification models (Individual Malware)

### C. *Comparison with Existing Works*

In terms of malware family classification, three of the four implemented classifiers produced better results than prior works regarding the CIC-Malmem2022 dataset, as seen in Table V. Utilizing genetic algorithms for feature selection, two-step outlier handling, and properly balancing the dataset using SMOTE oversampling and SOM-US undersampling played an integral part in higher accuracy. While feature selection and hyperparameter tuning were responsible for both reducing the time and size of the model. Our LightGBM and Decision Tree models are not only optimized but also outperformed previous works in every metric. This satisfies our research objective. Meanwhile, our Random Forest model increased the accuracy, precision, recall, and F1-score metric the most (by 4%) from the previous state-of-the-art model by Cassel and Majd [12].

From Table VI, we can observe, in terms of individual malware classification, all of our proposed models were able to outperform KNN with the Firefly Algorithm model with Random Forest and LightGBM in particular outperforming by 9% and 6.5%, respectively. Meanwhile, our Random Forest and LightGBM classifier models were able to outperform the Gray Wolf Optimization approach taken by Abualhaj et al. [3].

Therefore, we were able to create a state-of-the-art model in both the malware family classification and individual malware classification on the same dataset, setting a foundation for future research to build upon.

### D. *Preprocessing Pipeline Evaluation*

To further understand the aforementioned reasons that contributed to the high performance, comparison table VII and

| Models and Techniques | Precision | Recall | F1-score | Accuracy |
|---|---|---|---|---|
| Random Forest (with GA, SOM-US, SMOTE) | 0.911 | 0.912 | 0.912 | 0.912 |
| LightGBM (with GA, SOM-US, SMOTE) | 0.891 | 0.891 | 0.891 | 0.891 |
| Decision Tree (with GA, SOM-US, SMOTE) | 0.888 | 0.888 | 0.888 | 0.888 |
| KNN (with GA, SOM-US, SMOTE) | 0.852 | 0.852 | 0.852 | 0.852 |
| Random Forest (with SMOTE and TomeLink) [12] | 0.871 | 0.871 | 0.871 | 0.871 |
| Random Forest (with Gray Wolf Optimization) [3] | | | | 0.863 |
| Federated learning LSTM [18] | | | | 0.845 |
| KNN (with FireFly Optimization Algorithm) [10] | 0.838 | 0.837 | 0.836 | 0.837 |

TABLE V: Comparison Table with previous works (Malware Family)

| Models and Techniques | Precision | Recall | F1-score | Accuracy |
|---|---|---|---|---|
| Random Forest (with GA, SOM-US, SMOTE) | 0.788 | 0.787 | 0.787 | 0.787 |
| LightGBM (with GA, SOM-US, SMOTE) | 0.761 | 0.760 | 0.760 | 0.760 |
| Decision Tree (with GA, SOM-US, SMOTE) | 0.744 | 0.743 | 0.743 | 0.743 |
| KNN (with GA, SOM-US, SMOTE) | 0.710 | 0.709 | 0.709 | 0.710 |
| Random Forest (with Gray Wolf Optimization) [3] | 0.755 | 0.756 | 0.753 | 0.756 |
| KNN (with FireFly Algorithm) [10] | 0.714 | 0.696 | 0.700 | 0.696 |

TABLE VI: Comparison Table with previous works (Individual Malware)

Table VIII illustrate the results of family and individual classification using random forest before and after taking the key preprocessing steps.

| Metrics | Phase 1 | Phase 2 | Phase 3 |
|---|---|---|---|
| Accuracy | 0.872 | 0.911 | 0.912 |
| Precision | 0.872 | 0.911 | 0.911 |
| Recall | 0.872 | 0.911 | 0.912 |
| F1-score | 0.872 | 0.911 | 0.912 |
| Training Time | 16.70 s | 13.47 s | 8.2 s |
| Testing Time | 0.2610 s | 0.1678 s | 0.07 s |
| Model Size | 94.01 MB | 78.22 MB | 34.48 MB |

TABLE VII: Malware Family classification result using RF model

| Metrics | Phase 1 | Phase 2 | Phase 3 |
|---|---|---|---|
| Accuracy | 0.753 | 0.782 | 0.787 |
| Precision | 0.752 | 0.781 | 0.788 |
| Recall | 0.754 | 0.782 | 0.787 |
| F1-score | 0.752 | 0.781 | 0.787 |
| Training Time | 14.92 s | 8.69 s | 24.9 s |
| Testing Time | 0.337 s | 0.266 s | 0.201 s |
| Model Size | 315.85 MB | 313.95 MB | 234.71 MB |

TABLE VIII: Individual Malware classification result using RF model

**Phase 1:** Initial Results.
**Phase 2:** After incorporating 2-step class-wise outlier handling and feature selection.
**Phase 3:** After further incorporating dataset balancing and hyperparameter tuning.

Comparing phase 1 and phase 2 of the tables illustrates how the 2-step class-wise outlier handling and feature selection through genetic algorithm were integral to the high classification result while also contributing to lowering the train and test time and model size due to reduced dataset size. In phase 3, through undersampling and oversampling, a balanced dataset was created that maintained the accuracy, while hyperparameter tuning lowered the n_estimators and modifying some other parameters that resulted in further compression of the model and less training and testing time. One exception is the training time required for the individual malware classification. In phase 3, it increases from 8.69 seconds to 24.9 seconds. This increase is due to the dataset balancing, which ultimately increases the training sample. But in exchange, higher accuracy was achieved. Due to hyperparameter tuning, a lower model size was still maintained. Moreover, the test time reduces from phase 2, which is more essential to real-life deployment scenarios. Therefore, a comprehensive understanding of key contributors of the classification models for both the family and individual classifications can be observed.

## VI. Conclusion

### A. Concluding Remarks

The rapid increase of digital devices, particularly Internet of Things (IoT) devices, signifies the importance of protection from cyberattacks. These cyberattacks are ever-evolving, resulting in traditional signature-based services being unable to adapt to them. Therefore, approaches involving artificial intelligence become necessary. Although AI models already exist for modern-day computers and other high-end devices, there is a notable lack of models suitable for devices with lower computational capabilities. While the proposed preprocessing pipeline, alongside Random Forest and LightGBM classifiers, is capable of operating in such environments. The preprocessing pipeline, which includes class-based outlier handling and feature selection through Genetic Algorithm, improved upon the existing works and achieved greater accuracy than existing research in both malware family and individual malware classification. Meanwhile, hyperparameter tuning and effective data balancing through SMOTE and SOM-US contributed to optimization, resulting in a compact and efficient model suitable for resource-constrained devices.

### *B. Future Improvements*

While promising outcomes were achieved, there are doors to further research and refinement. The model can be improved by implementing optimized hybrid classification models and behavior-based anomaly detection models. This field of classification models is suitable for zero-day attacks. Zero-day attacks are those that take advantage of vulnerabilities of the system that are unknown to the vendor or developer. Incorporating that into the pre-existing model will allow for a robust system that is capable of dealing with the ever-evolving malware attack landscape and taking necessary steps to neutralize it by predicting the exact malware family or individual malware themselves. Another area of improvement would be to implement a fallback mechanism that would allow the model to work with a reduced set of features if it doesn't have the necessary tools to complete the full extraction process. Therefore, dynamically adjusting the input features based on available resources. This will allow for an adaptable malware detection process, suitable for a range of IoT devices with differing resource capabilities.